\RequirePackage{ifpdf}
\ifpdf 
\documentclass[pdftex]{sigma}
\else
\documentclass{sigma}
\fi

\begin{document}

\renewcommand{\PaperNumber}{008}

\FirstPageHeading

\ShortArticleName{2D Superintegrable Systems with 1-Parameter Potentials}

\ArticleName{Structure Theory for Second  Order\\ 2D Superintegrable Systems\\ with 1-Parameter Potentials}

\Author{Ernest G.~KALNINS~$^\dag$, Jonathan M.~KRESS~$^\ddag$, Willard MILLER Jr.~$^\S$ and Sarah POST~$^\S$}

\AuthorNameForHeading{E.G.~Kalnins, J.M.~Kress, W.~Miller Jr.\ and S.~Post}

\Address{$^\dag$~Department of Mathematics, University
of Waikato, Hamilton, New Zealand}
\EmailD{\href{mailto:math0236@math.waikato.ac.nz}{math0236@math.waikato.ac.nz}}
\URLaddressD{\url{http://www.math.waikato.ac.nz}}
\Address{$^\ddag$~School of Mathematics, The University of New South Wales,
Sydney NSW 2052, Australia}
\EmailD{\href{mailto:j.kress@unsw.edu.au}{j.kress@unsw.edu.au}}
\URLaddressD{\url{http://web.maths.unsw.edu.au/~jonathan/}}

\Address{$^\S$~School of Mathematics, University of Minnesota,
 Minneapolis, Minnesota,
55455, USA}
\EmailD{\href{mailto:miller@ima.umn.edu}{miller@ima.umn.edu},  \href{mailto:postx052@math.umn.edu}{postx052@math.umn.edu}}
\URLaddressD{\url{http://www.ima.umn.edu/~miller/}}

\ArticleDates{Received November 26, 2008, in f\/inal form January 14,
2009; Published online January 20, 2009}

\Abstract{The structure theory for the quadratic algebra  generated by f\/irst and second order constants of the motion for 2D second order superintegrable systems with nondegene\-ra\-te (3-parameter) and or 2-parameter  potentials is well understood,  but the results for the strictly 1-parameter case have been incomplete. Here we work out this structure theory and prove that the quadratic algebra generated by f\/irst and second order constants of the motion for systems with 4 second order constants of the motion must close at order three with the functional relationship between the  4 generators of order four. We also show that every  1-parameter superintegrable system is St\"ackel equivalent to a system on a constant curvature space.}


\Classification{20C99; 20C35; 22E70}

\section{Introduction}

A classical second order superintegrable system ${\cal H} =\sum_{ij}g^{ij}p_ip_j+V({\bf x})$ on an $n$-dimensional local Riemannian manifold is one that admits $2n-1$ functionally independent
symmetries ${\cal L}_k({\bf x}\cdot {\bf p} )$, $k=1,\dots,2n-1$ with ${\cal L}_1={\cal H}$, that are at most second order polynomials in the momenta $p_i$. (Further, at least one ${\cal L}_h=\sigma a^{ij}_h({\bf x})p_ip_j+W_h({\bf x})$ with $h>1$ must be exactly second order.) That is, $\{{\cal H},{\cal L}_k\}=0$ where
\[\{f,g\}=\sum_{j=1}^n(\partial_{x_j}f\partial_{p_j}g-\partial_{p_j}f\partial_{x_j}g)
\]
is the Poisson bracket for functions $f({\bf x},{\bf p})$, $g({\bf x},{\bf p})$, Here $2n-1$ is the maximum possible number of such symmetries, For the case $n=2$ The structure of the Poisson algebra generated by the symmetries has been the subject of great current interest. For potentials depending non-trivially on 2 or 3 parameters, see \cite{KKM20041}, for a precise def\/inition, it has been shown that the algebra is f\/inite-dimensional and closes at order six in the momenta. All such algebras have been classif\/ied, as have been all spaces and potentials that give rise to them \cite{KKM20042, DASK2005}. Similarly all degenerate 1-parameter potential systems are known via case-by case classif\/ication as well as the associated Poisson algebras, \cite{KKMP}. The number of true 1-parameter,  not just a restriction of a 3-parameter potential,  systems is~15 (6 in complex f\/lat space, 3 on the complex 2-sphere and one for each of the 4 Darboux spaces). Under the St\"ackel transform that maps superintegrable systems into equivalent systems on other manifolds, these divide into 6 equivalence classes. However information about the structures of the corresponding  algebras are known only by a~case by case listing and the mechanisms by which they close have never been worked out. Some results for 1-parameter potentials were reported in \cite{KKM20041} but although the results are correct the 1-parameter proofs are incomplete. Here we work out the structure theory and  prove that the quadratic algebra generated by f\/irst and second order constants of the motion for systems with 4 second order constants of the motion must close at order three and must contain a Killing vector. Furthermore we show that there must be a polynomial relation among the symmetries at order four. It is important to develop methods for understanding these structures that can be extended to structures for $n>2$ where the analysis becomes more complicated, and this approach should point the way.

We treat only classical superintegrable systems here, though the corresponding (virtually identical) results for the quantum systems follow easily \cite{KKM20061}. In both the classical and quantum cases the symmetry algebras and their representations have independent interest \cite{WOJ,EVA,EVAN,FMSUW,FSUW,BDK,CDas,GPS,KKW,KMPost3}. Also, there are deep connections with special functions and orthogonal polynomials, in particular Wilson polynomials \cite{KMJP,KMPost1}.

As an example, consider the classical Hamiltonian on the two
sphere
\[
{\cal H}={\cal J}^2_1+{\cal J}^2_2+{\cal J}^2_3+ \frac{a_3}{ s_3^2},
\]
where the ${\cal J}_i$ are def\/ined by ${\cal J}_3{=}s_1p_{s_2}{-}s_2p_{s_1}$
and cyclic permutation of indices, and \mbox{$s_1^2{+}s_2^2{+}s_3^2{=}1$}. If
we seek all f\/irst and second order constants of the motion for this classical
Hamiltonian, we f\/ind three possibilities in addition to $\cal H$ itself viz.
\[
{\cal A}_1={\cal J}^2_1+\frac{a_3 }{ 2s_3^2}\big(1+s_2^2-s_1^2\big),\qquad
{\cal A}_2={\cal J}_1{\cal J}_2-\frac{a_3 s_1s_2}{s_3^2},\qquad {\cal X}={\cal J}_3.\]
The set of 4 symmetries  ${\cal X}^2$, ${\cal H}$, ${\cal A}_1$ and ${\cal }A_2$ is linearly independent, but functionally dependent  via the fourth order identity
\[{\cal A}_1\big({\cal H}-{\cal A}_1-{\cal X}^2\big)-{\cal A}^2_2-\frac{a_3}{ 2}\big({\cal X}^2+{\cal H}\big)+\frac{a_3 ^2}{ 4}=0,\]
They satisf\/ies the Poisson algebra relations
\begin{gather*}
\label{classical1}
\{{\cal X},{\cal A}_1\}=-2{\cal A}_2,\qquad \{{\cal X},{\cal A}_2\}=-{\cal H}+{\cal X}^2+2{\cal A}_1,
\qquad  \{{\cal A}_1,{\cal A}_2\}=-{\cal X}(2{\cal A}_1+a_3 ),
\end{gather*}
so the algebra closes at order three. We will show that this structure is typical for all 1-parameter potentials that are not just restrictions of 3-parameter potentials, that is  there is always a Killing vector (a f\/irst order constant of the motion), the algebra always closes at order three, and there is always a fourth order relation between the 4 generators.

The situation changes drastically for the two sphere Hamiltonian with nondegenerate potential
\[V=\frac{a_1}{s_1^2}+\frac{a_2}{s_2^2}+\frac{a_3}{s_3^2},\]
where $s_1^2+s_2^2+s_3^2=1$.
The classical system has a basis of symmetries
\begin{gather*}
{\cal L}_1={\cal
  J}_1^2+a_2\frac{s_3^2}{s_2^2}+a_3\frac{s_2^2}{s_3^2},\qquad {\cal L}_2={\cal
  J}_2^2+a_3\frac{s_1^2}{s_3^2}+a_1\frac{s_3^2}{s_1^2},\qquad
{\cal L}_3={\cal
  J}_3^2+a_1\frac{s_2^2}{s_1^2}+a_2\frac{s_1^2}{s_2^2},
\end{gather*}
where ${\cal H}={\cal L}_1+{\cal L}_2+{\cal L}_3+a_1+a_2+a_3$
and the ${\cal J}_i$ are def\/ined by ${\cal J}_3=s_1p_{s_2}-s_2p_{s_1}$
and cyclic permutation of indices. The classical structure relations
are
\begin{gather*}
\{{\cal L}_1,{\cal R}\}=
8{\cal L}_1({\cal H}+a_1+a_2+a_3)-8{\cal
  L}_1^2-16{\cal L}_1{\cal L}_2 \nonumber\\
\phantom{\{{\cal L}_1,{\cal R}\}= }{} -16a_2{\cal L}_2+16a_3({\cal
  H}+a_1+a_2+a_3-{\cal L}_1-{\cal L}_2),
\\
\{{\cal L}_2,{\cal R}\}=
-8{\cal L}_2({\cal H}+a_1+a_2+a_3)+8{\cal
  L}_2^2+16{\cal L}_1{\cal L}_2\nonumber\\
  \phantom{\{{\cal L}_2,{\cal R}\}= }{} +16a_1{\cal L}_1-16a_3({\cal
  H}+a_1+a_2+a_3-{\cal L}_1-{\cal L}_2),\nonumber
\end{gather*}
with $\{{\cal L}_1,{\cal L}_2\}={\cal R}$ and
\begin{gather*}
{\cal R}^2-16{\cal L}_1{\cal L}_2({\cal H}+a_1+a_2+a_3) +16{\cal
  L}_1^2{\cal L}_2+16{\cal L}_1{\cal L}_2^2+16a_1{\cal
  L}_1^2+16a_2{\cal L}_2^2\\
\qquad{} +16a_3({\cal H}+a_1+a_2+a_3)^2
-32a_3({\cal
  H}+a_1+a_2+a_3)({\cal L}_1+{\cal L}_2)+16a_3{\cal L}_1^2\\
\qquad{} +32a_3{\cal
  L}_1{\cal L}_2 +16a_3{\cal L}_2^2-64a_1a_2a_3=0.
\end{gather*}
 Now there is no longer a f\/irst order symmetry but 3 second order symmetries.
 The
algebra generated by these symmetries and their commutators now
closes at order~6, \cite{KMPost2}.  The commutator~$\cal R$ cannot be expressible as a polynomial in the generators, but ${\cal R}^2$ and commutators of $\cal R$ with a generator can be so expressed. This 3-parameter system is called nondegenerate. Note that our 1-parameter potential is a restriction of the nondegenerate potential, but that the structure of the symmetry algebra has changed drastically.

On the other hand, the system with the 1-parameter potential
\[V=\frac{a}{s_1^2}+\frac{a}{s_2^2}+\frac{a}{s_3^2},\]
i.e., the restriction of the nondegenerate potential to the case $a=a_1=a_2=a_3$ has a symmetry algebra that is exactly the restriction of the algebra for the nondegenerate case. Further, any 2-parameter potentials obtained by restricting the nondegenerate potential can be shown not to introduce  symmetries in addition to those obtained by obvious restriction from the symmetry algebra for the nondegenerate case. We will show that these examples are typical for 2D second order superintegrable systems and will clarify the possible structures for the symmetry algebras in the degenerate cases.

\section{Background}
Before proceeding to the study of superintegrable systems with potential, we review some basic facts about second order symmetries  of the
underlying 2D complex Riemannian spaces. It is always possible to f\/ind a local  coordinate system $(x,y)\equiv(x_1,x_2)$
def\/ined in a neighborhood of $(0,0)$ on the manifold such that the metric takes the form
\[
ds^2=\lambda(x,y)(dx^2+dy^2)=\lambda\ dz\ d{\overline z},\qquad z=x+iy,\qquad {\overline z}=x-iy,
\]
and the Hamiltonian is ${\cal H}=(p_1^2+p_2^2)/\lambda+V(x,y)={\cal H}_0+V$, where $V$ is the potential function. We can consider a second order symmetry (constant of the motion)  as a quadratic form  ${\cal L}=\sum\limits_{i,j=1}^2a^{ij}(x,y)p_ip_j+W(x,y)$, $a^{ij}=a^{ji},$  that is in
involution with the  Hamiltonian $\cal H$: $\{{\cal H},{\cal L}\}=0$. A second order Killing tensor ${\cal L}_0=\sum\limits_{i,j=1}^2a^{ij}(x,y)p_ip_j$ is a symmetry of the free Hamiltonian ${\cal H}_0$: $\{{\cal H}_0,{\cal L}_0\}=0$. The Killing tensor conditions are
\begin{gather}a^{ii}_i=-\frac{\lambda_1}{\lambda}a^{i1} -\frac{\lambda_2}{\lambda}a^{i2},\qquad i=1,2;\nonumber\\
2a^{ij}_i+a^{ii}_j=-\frac{\lambda_1}{\lambda}a^{j1} -\frac{\lambda_2}{\lambda}a^{j2},\qquad i,j=1,2,\ i\ne j.
\label{Killingeqns}
\end{gather}
From these conditions we easily obtain the requirements
\[ 2a^{12}_1=-\big(a^{11}-a^{22}\big)_2,\qquad 2a^{12}_2=\big(a^{11}-a^{22}\big)_1.\]
{} From the integrability conditions for these last equations we see that
\[ \Delta a^{12}=0,\qquad \Delta \big(a^{11}-a^{22}\big)=0,\qquad \Delta=\partial_x^2+\partial_y^2.\]

In order for a form ${\cal L}={\cal L}_0+W$ to be a symmetry of the system ${\cal H}={\cal H}_0+V$ it is necessary and suf\/f\/icient that ${\cal L}_0$ be a Killing tensor and that $W$ satisfy the equation
\[ \{ {\cal H}_0,W\}+\{V,{\cal L}_0\}=0.\] The conditions for this are
\[ W_i=\sum_{j=1}^2a^{ij}V_j,\qquad i=1,2,\]
where $W_i=\partial_{x_i}W$, $V_j=\partial_{x_j}V$.  Necessary and suf\/f\/icient that these last two equations can be solved is the Bertrand--Darboux condition
\begin{gather}\label{BDcond} (V_{22}-V_{11})a^{12}+V_{12}\big(a^{11}-a^{22}\big)=
\left[\frac{\big(\lambda a^{12}\big)_1-\big(\lambda a^{11}\big)_2}{\lambda}\right]V_1+\left[\frac{\big(\lambda a^{22}\big)_1-\big(\lambda a^{12}\big)_2}{\lambda}\right]V_2.\end{gather}

For a second order superintegrable system we demand that there is a Hamiltonian and 2 other second order symmetries: ${\cal H}$, ${\cal L}_1$, ${\cal L}_2$ such that the Killing tensor parts of the 3 symmetries are functionally independent quadratic forms. By a change of basis if necessary, we can always assume that ${\cal L}_1$ is in Liouville form, so that the coordinates $x,y$ associated with ${\cal H}$, ${\cal L}_1$ are separable. Thus,
 we can choose our orthogonal coordinates $x$, $y=x_1$, $x_2$ such that the quadratic form in ${\cal L}_1$ satisf\/ies $a^{12}\equiv 0$, $a^{22}-a^{11}=1$. In this system
we have $\lambda_{12}=0$. A second symmetry is def\/ined by the Hamiltonian itself: $a^{11}=a^{22}=1/\lambda$, $a^{12}=0$,
which clearly always satisf\/ies
equations (\ref{Killingeqns}).
Due to functional independence, for the third symmetry ${\cal L}_2$ we must have $a^{12}\ne
0$ and it is on this
third symmetry that we will focus our attention  in the following.
Now the  integrability conditions can be rewritten as
\begin{gather}
\label {fundintcond1} \lambda_{12}=0,\qquad \Lambda\equiv \lambda_{22}-\lambda_{11}-3\lambda_1A_1+3\lambda_2A_2-\big(A_{11}+A_1^2-A_{22}-A_2^2\big)\lambda=0,
\end{gather}
where $A=\ln a^{12}$, the subscripts denote dif\/ferentiation and $A$ satisf\/ies $A_{11}+A_{22}+A_1^2+A_2^2=0$.
Equivalently,
\begin{gather}
\label{fundintcond2} \lambda_{12}=0,\ a^{12}_{11}+a^{12}_{22}=0,\qquad a^{12}(\lambda_{11}-\lambda_{22})+3\lambda_1a^{12}_1-3\lambda_2a^{12}_2+\big(a^{12}_{11}-a^{12}_{22}\big)\lambda=0.
\end{gather}
 In this second form a fundamental duality becomes evident \cite{Koenigs,KKM20041} (with a typo in the second reference): If $\lambda(x,y)$, $a^{12}(x,y)$ satisfy (\ref{fundintcond2}) then
\[{\tilde\lambda(x,y)}=a^{12}\left(\frac{x+iy}{\sqrt{2}},  \frac{-ix-y}{\sqrt{2}}\right),\qquad
{\tilde a}^{12}(x,y)=\lambda\left(\frac{x+iy}{\sqrt{2}},
\frac{-ix-y}{\sqrt{2}}\right)\]
also satisfy these conditions. Thus, the roles of metric and symmetry can be interchanged, and a second interchange returns the system to its original state.

Another  key equation  is the integrability condition  derived from consideration of $\Lambda_{12}=0$:
\begin{gather}\label{newintcondb0} 5L^{(1)}\lambda_1-5L^{(2)}\lambda_2+\big(L^{(1)}_1-L^{(2)}_2+3A_1L^{(1)}-3A_2L^{(2)}\big)\lambda=0,
\end{gather}
where $L^{(1)}=A_{112}-A_{12}A_1$, $L^{(2)}=A_{122}-A_{12}A_2$. We will derive this in detail  in Section \ref{section3}. The dual version of condition (\ref{newintcondb0}) is the integrability condition.
\begin{gather}\label{newintcondc0} 5K^{(1)}a^{12}_1-5K^{(2)}a^{12}_2+\big(K^{(1)}_1-K^{(2)}_2+3\rho_1K^{(1)}-3\rho_2K^{(2)}\big)a^{12}=0,
\end{gather}
where $\rho=\ln \lambda$ and $K^{(1)}=\rho_{222}-2\rho_{11}\rho_2-\rho_{22}\rho_2+\rho_1^2\rho_2$, $ K^{(2)}=-\rho_{111}+2\rho_{22}\rho_1+\rho_{11}\rho_1-\rho_2^2\rho_1$. Note that $K^{(1)}=K^{(2)}=0$ is the condition that $\lambda$ is a constant curvature space metric. Indeed, this is exactly the necessary and suf\/f\/icient condition that $\Delta (\ln \lambda)/\lambda=c$ where $c$ is a constant.

Koenigs \cite{Koenigs} employed condition  (\ref{newintcondc0}) to show that the only spaces admitting at least 6~li\-near\-ly independent constants of the motion were constant curvature spaces. Indeed in that case there are 3 functionally independent symmetries $a^{12}(x,y)$, hence 3~equations (\ref{newintcondc0}). That is only possible if the coef\/f\/icients of $a^{12}_1$, $a^{12}_2$ and $a^{12}$ vanish identically. Hence $K^{(1)}=K^{(2)}=0$ and $\lambda$~is a constant curvature  metric. Koenigs did not make use of condition (\ref{fundintcond1}), but from our point of view this condition is more fundamental.

Using our special coordinates and the Killing equations (\ref{Killingeqns}) we  can write the Bertrand--Darboux equations (\ref{BDcond}) in the form:
\begin{gather}\label{BDcond1} V_{12}=-\left[\frac{\lambda_2}{\lambda}\right]V_1-\left[\frac{\lambda_1}{\lambda}\right]V_2,\qquad V_{22}-V_{11}=\left[\frac{2\lambda_1}{\lambda}+3A_1\right]V_1-\left[\frac{2\lambda_2}{\lambda}+3A_2\right]V_2.
\end{gather}
In  \cite{KKM20041} we have shown that this system admits a nondegenerate i.e.\ 3-parameter (the maximum possible) potential $V(x,y)$ if and only if the potential is the general solution of the canonical system
\begin{gather}\label{3-parpot} V_{12}=A^{12}(x,y)V_1+B^{12}(x,y)V_2,\qquad V_{22}-V_{11}=A^{22}(x,y)V_1+B^{22}(x,y)V_2,
\end{gather}
whose integrability conditions are satisf\/ied identically. Thus this system will admit a 4-di\-men\-sio\-nal solution space, with one dimension corresponding to the trivial addition of an arbitrary constant. Each solution is uniquely determined at a point $(x_0,y_0)$ by prescribing the values $V$, $V_1$, $V_2$, $V_{11}$. In our special coordinates we have
\begin{gather}\label{3-parpoteqns} A^{12}=-\frac{\lambda_2}{\lambda},\qquad B^{12}=-\frac{\lambda_1}{\lambda},\qquad A^{22}=2\frac{\lambda_1}{\lambda}+3A_1,\qquad B^{22}=-2\frac{\lambda_2}{\lambda}-3A_2.
\end{gather}

We will say little about 2-parameter potentials other than pointing out, as we already showed in \cite{KKM20041} that they are just restrictions of 3-parameter potentials. Their canonical equations take the form
\begin{gather*}
V_{12}=A^{12}(x,y)V_1+B^{12}(x,y)V_2,\qquad V_{22}=A^{22}(x,y)V_1+B^{22}(x,y)V_2,
\\
V_{11}=A^{11}(x,y)V_1+B^{11}(x,y)V_2.
\nonumber
\end{gather*}

By relabeling coordinates, if necessary, we can always assume that the canonical equations for 1-parameter potentials take the form
\begin{gather*}
V_{1}=B^{1}(x,y)V_2,\qquad V_{22}-V_{11}=B^{22}(x,y)V_2,\qquad V_{12}=A^{11}B^{12}(x,y)V_2,
\end{gather*}
where the integrability conditions for these equations are satisf\/ied identically. This system will admit a 2-dimensional solution space, with one dimension corresponding to the trivial addition of an arbitrary constant. Each solution is uniquely determined at a point $(x_0,y_0)$ by prescribing the values $V$, $V_2$. In our special coordinates we have
\begin{gather}\label{1-parpoteqns} B^{12}=-\frac{\lambda_2}{\lambda}B^1-\frac{\lambda_1}{\lambda},\qquad B^{22}=\left(2\frac{\lambda_1}{\lambda}+3A_1\right)B^1-2\frac{\lambda_2}{\lambda}-3A_2.
\end{gather}

\section{The St\"ackel transform}
The importance of the St\"ackel transform in superintegrability theory is based on the following observation. Suppose we have a superintegrable system
\begin{gather*} 
{\cal H}=\frac{p_1^2+p_2^2}{\lambda(x,y)}+V(x,y)
\end{gather*}
in local orthogonal coordinates, with  $k$-parameter potential $V(x,y)$, $0\le k\le 3$
and suppose $U(x,y) $ is a particular choice of this potential for f\/ixed parameters, nonzero in an open set.  Then
the transformed system
\begin{gather*} 
{\tilde {\cal H}}=\frac{p_1^2+p_2^2}{{\tilde \lambda}(x,y)}+{\tilde V}(x,y),\qquad {\tilde \lambda}=\lambda U,\qquad  {\tilde V}=\frac{V}{U},
\end{gather*}
is also superintegrable.
Indeed, let ${\cal S}=\sum a^{ij}p_ip_j+W={\cal S}_0+W$  be a second order symmetry of $\cal H$ and ${\cal S}_U=\sum a^{ij}p_ip_j+W_U={\cal S}_0+W_U$  be the special case of this that is in involution with
$(p_1^2+p_2^2)/\lambda+ U$. Then it is straightforward to verify that
\begin{gather}\label{transformedsymmetry}{\tilde {\cal S}}={\cal S}_0-\frac{W_U}{U}H+\frac{1}{U}{\cal H}
\end{gather}
is the corresponding symmetry of $\tilde{\cal  H}$. Since one can always add a constant to a  potential, it follows that $1/U$ def\/ines an inverse St\"ackel transform of $\tilde {\cal H}$ to $\cal H$.
See \cite{HGDR,BKM, KKMW} for many examples of this transform. We say that two
superintegrable systems are {\it St\"ackel equivalent} if one can be
obtained from the other by a St\"ackel transform.  Note from (\ref{transformedsymmetry}) that the of\/f-diagonal elements $a^{12}=a^{21}$ of a symmetry remain invariant under the St\"ackel transform.

If $V(x,y)$ is a nondegenerate potential, i.e. the general (4-dimensional) solution of cano\-ni\-cal equations (\ref{3-parpot})
and $U(x,y)$ is a particular solution of these equations, then
 ${\tilde V}(x,y)$ is also a~nondegenerate potential satisfying the canonical equations
\begin{gather*}
{\tilde V}_{22}={\tilde V}_{11}+{\tilde A}^{22}{\tilde V}_1+{\tilde B}^{22}{\tilde V}_2,\qquad
{\tilde V}_{12}= {\tilde A}^{12}{\tilde V}_1+{\tilde B}^{12}{\tilde V}_2,
\end{gather*}
where
\[
{\tilde A}^{12}=A^{12}-\frac{U_2}{U},\qquad  {\tilde A}^{22}=A^{22}+2\frac{U_1}{U},\qquad  {\tilde B}^{12}=B^{12}-\frac{U_1}{U},\qquad  {\tilde B}^{22}=B^{22}-2\frac{U_2}{U}.
\]
Similarly, if $V(x,y)$ is a 1-parameter potential satisfying canonical equations
\begin{gather*}
V_{1}=B^{1}(x,y)V_2,\qquad V_{22}-V_{11}=B^{22}(x,y)V_2,\qquad V_{12}=B^{12}(x,y)V_2,
\end{gather*}
and $U(x,y)$ is a particular nonzero instance of this potential then ${\tilde V}(x,y)$ is also a  1-parameter  potential satisfying the canonical equations
\begin{gather*}
{\tilde V}_{1}={\tilde B}^{1}(x,y){\tilde V}_2,\qquad {\tilde V}_{22}-{\tilde V}_{11}={\tilde B}^{22}(x,y){\tilde V}_2,\qquad {\tilde V}_{12}={\tilde B}^{12}(x,y){\tilde V}_2,
\end{gather*}
where
\[
{\tilde B}^{12}=B^{12}-2B^{1}\frac{U_2}{U},\qquad  {\tilde B}^{22}=B^{22}+2\big(\big(B^1\big)^2-1\big)\frac{U_2}{U},\qquad  {\tilde B}^{1}=B^{1}.
\]
Note that the function $B^1$ remains invariant under a St\"ackel transform.

Now we return to study of a general superintegrable system with $k$-parameter potential, and viewed in our special coordinate system. Then $\lambda$ is the metric and $V$ is the general solution of the Bertrand--Darboux equations (\ref{BDcond1}). If the integrability conditions for these equations are satisf\/ied identically, then the solution space is $4$-dimensional. Otherwise the dimensionality is less. The metric $\lambda$ must satisfy the fundamental integrability conditions (\ref{fundintcond1}) that depend only on $a^{12}$, invariant under the St\"ackel transform. Now if $U$ is a particular solution of the Bertrand--Darboux equations (\ref{BDcond1}) then it def\/ines a St\"ackel transform to a new Riemannian space with metric $\mu=\lambda U$. Since $a^{12}$ is invariant under the transform the fundamental integrability conditions for $\mu$ are that same as for $\lambda$:
\begin{gather} \label{mueqns} \mu_{12}=0,\qquad \mu_{22}-\mu_{11}=3\mu_1A_1-3\mu_2A_2+\big(A_{11}+A_1^2-A_{22}-A_2^2\big)\mu.
\end{gather}
Note that these two equations  appear identical. However they have dif\/ferent interpretations. The f\/ixed metric $\lambda$ satisf\/ies (\ref{fundintcond1}) and is a special solution of
(\ref{mueqns}). Here $\mu$ designates  a $k+1$-dimensional family of solutions, of which $\lambda$ is a particular special case. It follows that $A$ satisf\/ies the
integrability conditions for this system. There is an isomorphism between the solutions $U$ of (\ref{BDcond1})  and the solutions $\mu=\lambda U$ of (\ref{fundintcond1}). The $\mu$ are precisely the metrics of all systems that can be obtained from the space with metric $\lambda$ via a St\"ackel transform.

\section{Review of known results}

For the most symmetric case, the potential is nondegenerate so $k=3$. Then the solution space of (\ref{mueqns}) is 4-dimensional. In \cite{KKM20042} we used this fact to derive all metrics and symmetries that correspond to superintegrable systems with nondegenerate potential.  The possible symmetries $A=\ln a^{12}$ that can appear are precisely the solutions of the Liouville equation $A_{12}=C e^A$ where $C$ is a constant. If $C=0$ then the system is equivalent via a St\"ackel transform to a~superintegrable system in f\/lat space. If $C\ne 0$ then the system is St\"ackel equivalent to the complex 2-sphere. As is easy to verify, these systems are precisely the solutions of the system of equations
\begin{gather*}
L^{(1)}\equiv A_{112}-A_{12}A_1=0,\qquad L^{(2)}\equiv A_{122}-A_{12}A_2=0.
\end{gather*}
An amazing fact is that these systems are exactly the same as those derived by Koenigs in his classif\/ication of all 2D spaces admitting at least 3 functionally independent second order Killing tensors.

A referee has called our attention to two recent and very interesting papers by Tsiga\-nov~\mbox{\cite{Tsiganov1, Tsiganov2}}.  He assumes that a superintegrable system admits an orthogonal  separation of variables in some coordinate system, so that there is a related St\"ackel matrix.  Under this assumption one can construct the action angle variables as explicit integrals. Using these Tsiganov shows a rough duality between superintegrability and functional addition theorems. which allows one to f\/ind a generating function for  constants of the motion, including those higher than second order. The Euler addition theorem for elliptic functions leads to the construction of all of the 2D superintegrable systems with nondegenerate potential! In this sense, these potentials are implicit in the work of Euler. This is clearly a powerful method for constructing superintegrable systems. It doesn't appear to yield any proof of completeness of the results or any classif\/ication of all possible spaces that admit superintegrability and in distinction to say \cite{{KKM20042}}  it requires the assumption of separation of variables.

 In \cite{KKM20041} we showed that all 2-parameter potentials were restrictions of nondegenerate potentials. Further, assuming the correctness of Koenigs' results we carried out a case by case analysis over many years to f\/ind all superintegrable systems with 1-parameter potentials. We found that all of these were restrictions of nondegenerate potentials again. However, in some cases the restricted potential admits a Killing vector, so that the structure of the associated quadratic algebra changes.  These results were  not based on a theoretical structure analysis. Our attempt at a structure analysis, contained in \cite{KKM20041} was incomplete and there were gaps in the proof, though the results are  correct.  Thus there is reason to use our St\"ackel transform approach to revisit this issue for 1-parameter potentials.

Finally the 0-parameter case deserves some attention, although it was already treated by Koenigs. If we consider a 0-parameter potential system as one in which it is not possible to admit a nonconstant potential, then it follows from our results and those of Koenigs that no such system exists. Koenigs' proof used complex variable techniques, very dif\/ferent from the methods used here.

\section{3-parameter and 2-parameter potentials}\label{section3}
Suppose we have a 2D second order superintegrable system with zero (or constant) potential, i.e., 3 functionally independent second order Killing tensors. Under what conditions does there exist a superintegrable system with nondegenerate potential $V$ such that the potential-free parts of the symmetries agree with the given Killing tensors? To answer this we choose the special coordinates
such that the integrability conditions for the zero potential case are:
\begin{gather*}
\lambda_{12}=0,\qquad \Lambda\equiv \lambda_{22}-\lambda_{11}-3\lambda_1A_1+3\lambda_2A_2-\big(A_{11}+A_1^2-A_{22}-A_2^2\big)\lambda=0,
\\
 \Delta\equiv A_{11}+A_{22}+A_1^2+A_2^2=0,\nonumber
 \end{gather*}
where $A=\ln a^{12}$.
Necessary and suf\/f\/icient conditions that this system admits a nondegenerate potential $V$ with canonical equations
\begin{gather}\label{veqn11a}
V_{22}=V_{11}+A^{22}V_1+B^{22}V_2,\qquad
V_{12}= A^{12}V_1+B^{12}V_2,
\end{gather}
where $A^{ij}$, $B^{ij}$ are given by (\ref{3-parpoteqns}), are that the integrability conditions for equations (\ref{veqn11a}) are satisf\/ied identically. These conditions are:
\begin{gather}\label{3potintconds}
T^{(1)}\equiv 2B^{12}_2-B^{22}_1-2A^{12}_x-A^{22}_2=0,
\\
 T^{(2)}\equiv  2B^{12}_2A^{22}-A^{22}B^{22}_1-A^{22}A^{12}_1-2A^{12}B^{12}_1-A^{22}_{12}+A^{12}_{22}+2A^{12}A^{12}_2\nonumber\\
\phantom{T^{(2)}\equiv}{}
+B^{12}A^{22}_2-B^{22}_2A^{12}-B^{22}A^{12}_2-A^{12}_{11}=0,\nonumber\\
  T^{(3)}\equiv -B^{12}A^{22}_1+2A^{12}_2B^{12}+B^{22}B^{12}_2-B^{22}B^{22}_1-2B^{12}B^{12}_1-A^{22}B^{12}_1\nonumber\\
\phantom{T^{(3)}\equiv}{} +A^{12}B^{22}_1+B^{12}_{22}-B^{22}_{12}-B^{12}_{11}=0.\nonumber
\end{gather}

Substituting expressions  (\ref{3-parpoteqns}) into (\ref{3potintconds}) we f\/ind that $T^{(1)}=0$ identically. To understand the remaining conditions we use $\Lambda=0$, $\Delta=0$ to simplify the equations. We solve for $\lambda_{111}$ and $\lambda_{222}$ from $\Lambda_1=0$, $\Lambda_2=0$, respectively, and substitute these expressions in $T^{(3)}$ and $T^{(2)}$, respectively. Then we solve for $\lambda_{11}-\lambda_{22}$ from $\Lambda=0$ and substitute this result into $T^{(3)}$ and $T^{(2)}$.  Then we f\/ind
\[T^{(3)}+\Delta_1=5L^{(2)},\qquad T^{(2)}-\Delta_2=-5L^{(1)},\]
or
\[ T^{(3)}=5L^{(2)}\quad {\rm mod }\ \Delta,\qquad T^{(2)}=-5L^{(1)}\quad {\rm mod}\ \Delta,\]
where we say $F=G \ {\rm mod}\ \Delta$ if $F-G$ is a functional linear combination of derivatives of $\Delta$.
Thus the superintegrable system admits a nondegenerate potential $V$ if and only if $L^{(1)}=L^{(2)}=0$. This last condition  exactly
characterizes the spaces classif\/ied by Koenigs: f\/lat space, the 2-sphere, the 4 Darboux spaces, and the family we call Koenigs spaces. Thus it is a consequence of Koenigs' classif\/ication is that all spaces admitting 3 second order Killing tensors automatically admit a nondegenerate potential.

In establishing the above result we have not made use of all of the information obtainable from the symmetry integrability conditions (\ref{mueqns}). In the  identity $\Lambda_{12}=0$ the second derivative terms in $\lambda$ appear in the form $A_{12}(\lambda_{22}-\lambda_{11})+ \cdots =0$. If $A_{12}=0$ then $L^{(1)}=L^{(2)}=0$ and we have one of the spaces found by Koenigs. Suppose $A_{12}\ne 0$. Then we have a new integrability condition
\[\Omega \equiv\left(\frac{\Lambda_{12}}{3A_{12}}\right)_{12}=0.\]
Solving for $\mu_{22}-\mu_{11}$ from $\Lambda=0$ and substituting into $\Omega=0$ we obtain, after a straightforward computation,  a condition of the form
\[ S^{(1)}\mu_1+S^{(2)}\mu_2+S\mu=0,\]
where
\begin{gather*}
S^{(1)}=5L^{(1)}\quad {\rm mod}\ \Delta,\qquad S^{(2)}=-5L^{(2)}\quad {\rm mod}\ \Delta,\\
S=L^{(1)}_1-L^{(2)}_2+3A_1L^{(1)}-3A_2L^{(2)} \quad {\rm mod}\ \Delta.
\end{gather*}
Thus we have the integrability condition
\begin{gather}\label{newintcond} 5L^{(1)}\mu_1-5L^{(2)}\mu_2+\big(L^{(1)}_1-L^{(2)}_2+3A_1L^{(1)}-3A_2L^{(2)}\big)\mu=0.
\end{gather}
When a nondegenerate potential exists then the space of solutions of (\ref{mueqns}) is 4-dimensional and the values of $\mu$, $\mu_1$, $\mu_2$, $\mu_{11}$ can be prescribed arbitrarily at any regular  point. Thus the integrability condition (\ref{newintcond}) can hold only if $L^{(1)}=L^{(2)}=0$. This proves that the systems admitting a~nondegenerate potential coincides with the potential free systems found by Koenigs. The same argument goes through for the case when a 2-parameter potential exists. Then the space of solutions of (\ref{mueqns}) is 3-dimensional and the values of $\mu$, $\mu_1$, $\mu_2$ can be prescribed arbitrarily at any regular  point, so the integrability condition (\ref{newintcond}) can hold only if $L^{(1)}=L^{(2)}=0$. This proves that any 2-parameter potential must be a restriction of a nondegenerate potential, a fact proved by a dif\/ferent method in~\cite{KKM20041}.

\section{1-parameter potentials} \label{1parsection}
The theory of 1-parameter potentials is more complicated than that for nondegenerate and 2-parameter potentials, due to the possible occurrence of systems with either 4 or 3  linearly independent second order symmetries. Suppose $V(x,y)$ is a 1-parameter potential satisfying canonical equations
\begin{gather*}
V_{1}=B^{1}(x,y)V_2,\qquad V_{22}-V_{11}=B^{22}(x,y)V_2,\qquad V_{12}=B^{12}(x,y)V_2.
\end{gather*}
The integrability conditions for these equations are
\begin{gather} B^{12}\big(1-\big(B^1\big)^2\big)-B^1_2-B^1B^1_1-B^1B^{22}=0,\nonumber\\
B^{12}_2-B^{22}_1-B^1_{11}-B^1_1B^{12}-B^1B^{12}_1=0.\label{1parintconds}
\end{gather}
In special coordinates  $B^{22}$, $B^{11}$ are given in terms of $B^1$ by relations~(\ref{1-parpoteqns}), so the f\/irst equa\-tion~(\ref{1parintconds}) becomes
\begin{gather}\label{1parintcond1} 3\big(A_2-B^1A_1\big)=\frac{B^1_2+B^1B^1_1}{B^1}+
\left(\frac{\lambda_1}{\lambda}-B^1\frac{\lambda_2}{\lambda}\right)\left(\frac{1}{B^1}+B^1\right),
\end{gather}
unless  $B^1\equiv 0$, in which case it becomes $B^{12}=\lambda_1=0$. Note also the special case $B^1\equiv \pm i$, which implies $A_2=A_1B^1$.

The second equation (\ref{1parintconds}) becomes
\begin{gather} \lambda(\lambda_{11}+\lambda_{22})B^1+\lambda_1\lambda_2\big(1+\big(B^1\big)^2\big)-\big(\lambda_1^2+\lambda_2^2\big)B^1
+\lambda\lambda_1B^1_1\nonumber\\
\qquad{}+\lambda\lambda_2\big(B^1_2-2B^1_1B^1\big)+\lambda^2\big(B^1_{11}+B^1_1A_1+3B^1A_{11}-3A_{12}\big)=0.
\label{1parintcond2}
\end{gather}

Substituting (\ref{1parintcond1}) into the expression (\ref{1-parpoteqns}) for $B^{22}$ we f\/ind
\begin{gather}\label{newB22} B^{22}=\frac{\big(\frac{\lambda_1}{\lambda}+\frac{\lambda_2}{\lambda}B^1\big)\big(\big(B^1\big)^2-1\big)
-\big(B_2^1+B^1B^1_1\big)}{B^1}.
\end{gather}

Now we write down the involutory system of equations that determine the second order symmetries of a 1-parameter potential system
\begin{gather}a^{11}_1=-\frac{\lambda_1}{\lambda}a^{11}-\frac{\lambda_2}{\lambda}a^{12},\nonumber\\
 a^{11}_2=-\frac{\lambda_2}{\lambda}a^{11}+\frac{\lambda_2}{\lambda}\big(a^{11}-a^{22}\big)
 -\frac{\lambda_1}{\lambda}a^{12}-2b,\nonumber\\
\big(a^{11}-a^{22}\big)_1=\frac23\left(-B^{22}+2\frac{\lambda_1}{\lambda}B^1
-2\frac{\lambda_2}{\lambda}\right)a^{12}+2B^1b,\nonumber\\
\big(a^{11}-a^{22}\big)_2=-2b,\nonumber\\
a^{12}_1=b,\nonumber\\
a^{12}_2=\frac13\left(-B^{22}+2\frac{\lambda_1}{\lambda}B^1-2\frac{\lambda_2}{\lambda}\right)a^{12}+B^1b,\nonumber\\
b_1= -\frac12 a^{11}_2 -\frac12 \partial_x\left( \frac{\lambda_1}{\lambda} a^{12}+\frac{\lambda_2}{\lambda} a^{22}\right),\nonumber\\
b_2= -2B^1b_1-2B^1_1b+\partial_y\left(\left[\frac13(2\frac{\lambda_1}{\lambda}B^1
-2\frac{\lambda_2}{\lambda}-B^{22}\right]a^{12}\right).\label{1parsymcond}
\end{gather}
There is one additional condition, obtained by dif\/ferentiating one of the Killing equations (\ref{Killingeqns}), that we have not made use of in obtaining the involutory system:
\begin{gather}\label{modifiedBD} 0=2a^{12}_{22}+a^{22}_{12}+\left(\frac{\lambda_1}{\lambda}a^{11}+\frac{\lambda_2}{\lambda}a^{12}\right)_2.
\end{gather}
When the indicated dif\/ferentiations and substitutions are carried out, the right hand side of each of these equations can be expressed in terms of the variables $a^{11}$, $a^{11}-a^{22}$, $a^{12}$, $b=a^{12}_1$ alone, although the expanded terms are lengthy. We think of $B^1$, $B^{22}$ as explicit given functions.

Note that though this system is in involution, the system without the added variable $b=a^{12}_1$ is not. This demonstrates that a symmetry is uniquely determined by the values of $a^{11}$,
$a^{22}$, $a^{12}$ and $a^{12}_1$ at a regular point; the values of $a^{11}$, $a^{22}$, $a^{12}$ may not suf\/f\/ice. By assumption, the system has 3 functionally independent second order symmetries. However, the involutory system indicates that there may, in fact, be 4~linearly independent second order symmetries, but obeying a functional dependence relation.

Now we require that the system (\ref{1parsymcond}), (\ref{modifiedBD}) admit 4 linearly independent solutions. Then at a~regular point there exists a unique solution with any prescribed values of $a^{11}$, $a^{22}$, $a^{12}$, $b$ and the integrability conditions for the system are satisf\/ied identically in these variables. To investigate the properties of this system we expand the condition (\ref{modifiedBD}) in terms of the basic 4~variables. The result takes the form $D^1(x,y)a^{12}+D^2(x,y)b=0$, where the $D^j$ do not depend on the basic variables. Indeed,
\[D^2=4B^1B^{22}-6\big(B^1B^1_1+B^1_2\big)-\lambda_2B^1
+9\frac{\lambda_1}{\lambda}+\frac{\lambda_1}{\lambda}\big(B^1\big)^2-9\frac{\lambda_2}{\lambda}\big(B^1\big)^3,\]
with a similar but more complicated formula for $D^1$. Since the variables can be prescribed arbitrarily at a regular point $(x,y)$, the only way for condition (\ref{modifiedBD}) to hold is for $D^1=D^2=0$. We solve for $B^1B^1_1+B^1_2$ from the equation $D^2=0$ and substitute this result into (\ref{newB22}) to obtain an updated expression for $B^{22}$ that is independent of $A^{12}$:
\begin{gather*}
B^{22}(x,y):=\frac12\left(\frac{-\lambda_2 B^1+\lambda_1\big(B^1\big)^2+3 \lambda_2\big(B^1\big)^3-3\lambda_1}{B^1\lambda}\right).
\end{gather*}
(Here we are assuming $B^1\ne 0$. This special case will be treated separately.)  We substitute this updated expression for $B^{22}$ in all of the previous equations, and eliminate $B^1_2$ from all expressions, including (\ref{1parintcond2}).  The condition $D^1=0$ is now satisf\/ied identically  with the updated expressions. The only remaining constraints are the integrability conditions for the symmetry equations (\ref{1parsymcond}). These conditions are satisf\/ied identically except for $\partial_1b_2=\partial_2b_1$ which takes the form $E^1(x,y)a^{12}+E^2(x,y)b=0$, where the $E^j$ do not depend on the basic variables. Thus $E^j(x,y)=0$ for the case where 4 linearly independent symmetries exist.

In the case where the space of symmetries is strictly 3-dimensional, the integrability conditions will no longer be satisf\/ied identically, since there is a linear condition satisfying by the variab\-les~$a^{11}$, $-a^{22}$, $a^{12}$, $b$  Thus, for example the condition $D^1(x,y)a^{12}+D^2(x,y)b=0$ should now be considered as a constraint relating $a^{12}$ and~$b$.

These integrability equations for the 4-dimensional and 3-dimensional cases are rather complicated and their geometric signif\/icance is not clear, so we will pass to a simpler, St\"ackel transform approach, while making use of the partial results we have obtained via the direct integrability condition attack.

\section{The St\"ackel transform for 1 parameter potentials\\ with 4 linearly independent symmetries}

To shed more light on this case we follow the approach of Section \ref{section3}. That is, we choose special coordinates and restrict our attention to the symmetries for which $a^{12}\ne 0$, essentially a two-dimensional vector space. Since $B^1$ and $a^{12}$ are invariant under St\"ackel transformations and the equations for the symmetries and the metric $\lambda$ depend only on these variables, these equations are identical for all metrics $\mu$ describing systems St\"ackel equivalent to the original one. The basic equations are the symmetry conditions
\begin{gather*} \mu_{12}=0,\qquad \Lambda\equiv \mu_{22}-\mu_{11}-3\mu_1A_1+3\mu_2A_2-\big(A_{11}+A_1^2-A_{22}-A_2^2\big)\mu=0,\nonumber
\\
\Delta\equiv A_{11}+A_{22}+A_1^2+A_2^2=0,
\end{gather*}
where $A=\ln a^{12}$,
and the integrability conditions for the 1-parameter  potential $V$, where $V_1=B^1V_2$. Writing $B^1=B$ for short and using the partial results obtained in the preceding section for simplif\/ication, we can write the f\/irst integrability condition (\ref{1parintcond1}) for the potential as
\begin{gather}\label{1parintcond1b} \mu_1-B\mu_2+ D\mu=0,\qquad { D}=3B\frac{BA_1-A_2}{B^2+1}+\frac{B_2+BB_1}{B^2+1},
\end{gather}
and the second integrability condition
(\ref{1parintcond2}) as
\begin{gather} \mu_{11}\mu\big(B^3+3B\big)-2\mu_{22}\mu B^3-\mu_1^2\big(B^3+3B\big)+\mu_2^2 \big(3B^3+B^5\big)\nonumber\\
\qquad{}
+\mu_1\mu\big(B^2B_1-3B_1\big)-2\mu^2B^2B_{11}=0.\label{1parintcond2b}
\end{gather}

These equations have a dif\/ferent interpretation than those of the last section. First of all they hold for 2 distinct functions $a^{12}$ whose ratio is nonconstant. Secondly, the space of solutions~$\mu$ of this system is 2-dimensional. Thus at any regular point $(x,y)$ there is a unique solution $\mu(x,y)$ taking on prescribed values $\mu$, $\mu_2$ at the point. Use of (\ref{1parintcond1b}) and dif\/ferentiation yields the linear expressions
\begin{gather}\label{muexp} \mu_1=B\mu_2-{D}\mu,\!\!\qquad  \mu_{11}=(B_1-DB)\mu_2+\big(D^2-D_1\big)\mu,\qquad\! \! \mu_{22}=\frac{D_2}{B}\mu+\frac{D-B_2}{B}\mu_2,\!\!\!\!
\end{gather}
for $\mu_1$, $\mu_{11}$, $\mu_{22}$ in terms of $\mu$ and $\mu_2$.

There are additional integrability and compatibility conditions for the system (\ref{1parintcond1b}), (\ref{1parintcond1b}) that constrain $A$ and $B$. The only nontrivial integrability condition for the subsystem $\mu_{12}=0$, $\mu_1-B\mu_2+ D\mu=0$, is
\[
\left [\left(\frac {D_2}{B}\right)_1-\frac{DD_2}{B}\right]\mu+\left[D_2+\left(\frac{D-B}{B}\right)_1\right]\mu_2=0.
 \]
Since this must hold for all solutions $\mu$ we have the 2 conditions
\begin{gather}\label{intcondsA} \left(\frac {D_2}{B}\right)_1-\frac{DD_2}{B}=0,\qquad D_2+\left(\frac{D-B_2}{B}\right)_1=0.
\end{gather}
The requirement that the subsystem be compatible with $\Lambda=0$ is
\[
\left(\frac{D_2}{B}-D^2+D_1+3A_1D-C\right)\mu+\left(\frac{D-B_2}{B}+DB-B_1-3A_1B+3A_2\right)\mu_2=0,
\]
i.e.,
\begin{gather}\label{compatcondsA}
\frac{D_2}{B}-D^2+D_1+3A_1D-C=0,\qquad \frac{D-B_2}{B}+DB-B_1-3A_1B+3A_2=0,
\end{gather}
where $C=A_{11}+A_1^2-A_{22}-A_2^2$. We already knew the second of conditions (\ref{compatcondsA}) but, from the analysis of the previous section,  the requirement of a 4-dimensional space of symmetries yields the identity
$B_2+BB_1=B(A_2-BA_1)$, so $D$ has the alternate expression
\begin{gather*}
D=-2\frac{B_2+BB_1}{B^2+1}.
\end{gather*}
 This is the full set of integrability conditions.

The key to understanding these systems is the f\/irst order condition (\ref{1parintcond1b}). We will show that this equation, together with other integrability conditions, implies that each  system in the family admits a Killing vector, which is also a f\/irst order symmetry of the system. To see this, consider the condition that the form ${\cal X}=\xi p_1+\eta p_2$ be a Killing vector. This condition is simply that the Poisson bracket of $\cal X$ and the free Hamiltonian ${\cal H}_0$ vanish, i.e.,
\[ \left\{ \xi p_1+\eta p_2,\frac{p_1^2+p_2^2}{\mu}\right\}=0.\]
Thus the coef\/f\/icients of $p_1^2$, $p_2^2$, $p_1p_2$ in the resulting expression must vanish:
\begin{gather*}
2\xi_1\mu+\xi\mu_1+\eta\mu_2=0,\qquad 2\eta_2\mu+\xi\mu_1+\eta\mu_2=0,\qquad \eta_1+\xi_2=0.
\end{gather*}
We f\/ind that
\[ \eta_1=-\xi_2,\qquad \xi_1=\eta_2,\qquad {\rm where}\qquad 2\xi_1\mu+\xi\mu_1+\eta\mu_2=0.
\]
In order that (\ref{1parintcond1b}) be interpretable as the Killing vector requirement there must exist an integra\-ting factor $Q$ such that
$Q\mu_1-QB\mu_2+ QD\mu=0$ and $Q=\xi$, $\eta=-QB$, $\xi_1=QD/2.$ Further we must require $QD/2=\eta_2=-Q_2B-QB_2=\xi_1=Q_1$, and $Q_2=\xi_2=-\eta_1=Q_1B+QB_1$. Thus we
obtain the system
\[
 (\ln Q)_1=D/2,\qquad (\ln Q)_2=B_1+BD/2
 \]
whose integrability condition is
\begin{gather}\label{TARGET} D_2=2B_{11}+B_1D+BD_1,\qquad {\rm or}\qquad \big(1+B^2\big)(B_{11}+B_{22})-2B\big(B_1^2+B_2^2\big)=0.
\end{gather}
This condition is a consequence of (\ref{1parintcond2b}) and the second integrability condition (\ref{intcondsA}). Indeed, using~(\ref{muexp}) to express $\mu_{11}$, $\mu_1$, $\mu_{22}$ in terms of $\mu_2$ and $\mu$ in (\ref{1parintcond2b}) we f\/ind that the resulting expression takes the form
$F(x,y)\mu^2=0$, so $F(x,y)=0$. Solving for $B_{12}$ in each of $F(x,y)=0$ and (\ref{intcondsA}), and equating the results, we get exactly the desired condition~(\ref{TARGET}).

We have shown that ${\cal X}=\xi p_1+\eta p_2=Qp_1-QBp_2$ is a Killing vector. Moreover $\cal X$ is a f\/irst order symmetry, since
\[ \{{\cal X},{\cal H}\}=\{{\cal X},{\cal H}_0\} +\{{\cal X},V\}=-Q(V_1-BV_2)=0.\]

Now we check the special cases $B=0,\pm i$. The cases $B=\pm i$ are essentially the same. Choosing $B=-i$, by interchanging $x$ and $y$ if necessary, we see that the second integrability condition for the potential gives the condition $\Delta \ln (\lambda)=0$, i.e., the condition that $\lambda$ is a~f\/lat space metric. Solving this equation, and identifying solutions that are equivalent under Euclidean transformations and dilations, we f\/ind three cases:
\[ {\rm I:} \ \lambda=1,\qquad {\rm II:} \ \lambda= e^y,\qquad {\rm III:} \ \lambda =x^2+y^2.\]
For the f\/irst case a straightforward computation yields the Killing vector ${\cal X}=p_2-ip_1$ and the potential $V=\alpha(y-ix)$. This is the superintegrable system $[E4]$ in~\cite{KKMP}.
For case II the Killing vector is ${\cal X}=e^{-(y+ix)/2}(p_2-ip_2)$  and the potential is $V=\alpha e^{-(y-ix)}$ (corresponding to the superintegrable system $[E14]$ in \cite{KKMP}). For case~III the Killing vector is $p_2-ip_1$ and the potential is $V=1/(y-ix)$ (corresponding to the superintegrable system $[E13]$ in~\cite{KKMP}). In all these cases, $A_{12}=0$. If $B=0$ then $\lambda_1=0$ and the Killing vector is $p_1$. Thus $V=V(y)$ and all metrics $\mu$ St\"ackel equivalent to $\lambda$ will satisfy $\mu_1=0$, so the fundamental equations are
\[\mu_{1}=0,\qquad  \Lambda\equiv \Lambda\equiv \mu_{22}+3\mu_2A_2-\big(A_{11}+A_1^2-A_{22}-A_2^2\big)\mu=0.\]
These equations must have a 2-dimensional vector space of solutions, so that $\mu$ and $\mu_2$ can be prescribed arbitrarily at a regular point. Since
$\Lambda_1\equiv 3\mu_2A_{12}-{\cal C}_1\mu=0$ where ${\cal C}=A_{11}+A_1^2-A_{22}-A_2^2$, this implies $A_{12}={\cal C}_1=0$. Thus the spaces are contained in our earlier classif\/ication.

\begin{theorem}\label{theorem2}If a $2D$ superintegrable system with a $1$-parameter potential admits $4$ linearly independent second order symmetries, then it also admits a Killing vector. One of the second order symmetries is the square of the Killing vector.
\end{theorem}

Now we return to the generic  case where $B\ne 0,\pm i$.  We  show that $B(x,y)=B^1$ always factors, so that $V_1=X(x)Y(y)V_2$.
\begin{lemma} \label{Bfactors}
\[
B_{12}=\frac{B_1B_2}{B}\qquad {\rm so}\qquad (\ln B)_{12}=0 \qquad {\rm and}\qquad  B=X(x)Y(y).
\]
\end{lemma}

\begin{proof}
We solve for $B_{12}\equiv B^1_{12}$ from (\ref{1parintcond2b}), i.e., from $F(x,y)=0$. Then we solve for $B_{11}$ from~(\ref{TARGET}) and substitute this result into our expression for $B_{12}$. We obtain $B_{12}=B_1B_2/B$. \end{proof}

 Since ${\cal X}=Qp_1-QBp_2$ is a Killing vector, ${\cal X}^2$ is a second order symmetry with $-a^{12}=Q^2 B$. Thus $A=\ln a^{12}=2\ln Q
+\ln B$ where  $(\ln Q)_1=D/2$, $(\ln Q)_2=B_1+BD/2$. From this it is straightforward to compute  the derivatives of $A$ in terms of $B$ and its derivatives. We have
\begin{gather*}
A_1=-2\frac{B_2+BB_1}{B^2+1}+\frac{B_1}{B},\qquad A_2=2B_1-2\frac{(B_2+BB_1)B}{B^2+1}+\frac{B_2}{B},\\
A_{12}=\frac{2B_{22} +B_2B_1+B_{12}}{B^2+1}+\frac{4(B_2+BB_1)BB_2}{(B^2+1)^2} +\frac{B_{12}B-B_2B_1}{B^2},
\end{gather*}
with analogous expressions for $A_{112}$ and $A_{122}$.  Next we substitute $B=X(x)Y(y)$ into each of these expressions, and in the identity (\ref{TARGET}), and then compute $L^{(1)}=A_{112}-A_{12}A_1$ and $L^{(2)}=A_{122}-A_{12}A_2$ in terms of $X(x)$, $Y(y)$. Solving for $X''(x)$ from the identity (\ref{TARGET}) and substituting this expression into $L^{(1)}$, $L^{(2)}$, we f\/ind
\[L^{(1)}=L^{(2)}=0.\]

\begin{theorem}\label{L1L2=0} If $2D$ superintegrable system with a $1$-parameter potential $\alpha V$ admits $4$~linearly independent second order symmetries then there exists a superintegrable system with nondege\-ne\-rate potential $\tilde V(\alpha,\beta,\gamma)$ such that the restriction  $\tilde V(\alpha,0,0)=\alpha V$ and  the restricted second order symmetries of the nondegenerate system agree with a three-dimensional subspace of the second order symmetries for the $1$-parameter potential.
\end{theorem}

For future use we note that the two-dimensional space of nonzero symmetries $a^{12}$ (excluding the zero function) does not necessarily have the property that  $a^{12}$ satisf\/ies the Liouville equation $\Delta (\ln a^{12})=ca^{12}$. However, in the generic case $B\ne 0,\pm i$ we have shown that there is a Killing vector ${\cal X}=\xi p_1+ \eta p_2$ and that the associated $a^{12}=\xi\eta$ from the symmetry ${\cal X}^2$ is nonzero and does satisfy the Liouville equation.

\section{The St\"ackel transform for 1-parameter potentials\\ with exactly three linearly independent symmetries}

Again we follow the approach of Section~\ref{section3} and restrict our attention to the symmetries for which $a^{12}\ne 0$, now a one-dimensional vector space. Since $B^1$ and $a^{12}$ are invariant under St\"ackel transformations and the equations for the symmetries and the metric $\lambda$ depend only on these variables, these equations are identical for all metrics $\mu$ describing systems St\"ackel equivalent to the original one. The basic equations are the symmetry conditions
\begin{gather*}
\mu_{12}=0,\qquad \Lambda\equiv \mu_{22}-\mu_{11}-3\mu_1A_1+3\mu_2A_2-\big(A_{11}+A_1^2-A_{22}-A_2^2\big)\mu=0,
\\
 \Delta\equiv A_{11}+A_{22}+A_1^2+A_2^2=0,
 \nonumber
\end{gather*}
where $A=\ln a^{12}$,
and the integrability conditions for the 1-parameter  potential $V=V(x,y)$:
\begin{gather*}
V_{1}=B^{1}(x,y)V_2,\qquad V_{22}-V_{11}=B^{22}(x,y)V_2,\qquad V_{12}=B^{12}(x,y)V_2.
\end{gather*}
The integrability conditions for these equations are
\begin{gather} B^{12}\big(1-\big(B^1\big)^2\big)-B^1_2-B^1B^1_1-B^1B^{22}=0,\nonumber\\ B^{12}_2-B^{22}_1-B^1_{11}-B^1_1B^{12}-B^1B^{12}_1=0.\label{1parintconds1}
\end{gather}
In a special coordinate system $B^{22}$, $B^{11}$ are given in terms of $B^1$ by relations (\ref{1-parpoteqns}), so the f\/irst equation
(\ref{1parintconds1}) becomes
\begin{gather}\label{1parintcond1a} \mu_1-B^1\mu_2+D\mu=0,\qquad D=\frac{B^1_2+B^1B^1_1-3B^1(A_2-B^1A_1)}{1+(B^1)^2}.
\end{gather}
(The cases $B^1\equiv 0,\pm i$ cannot occur because we have seen that they lead to 4 independent symmetries.) The second equation becomes
\begin{gather*}
\mu(\mu_{11}+\mu_{22})B^1+\mu_1\mu_2\big(1+\big(B^1\big)^2\big)-\big(\mu_1^2+\mu_2^2\big)B^1+\mu\mu_1B^1_1\nonumber
\\
\qquad {} +\mu\mu_2\big(B^1_2-2B^1_1B^1\big)+\mu^2\big(B^1_{11}+B^1_1A_1+3B^1A_{11}-3A_{12}\big)=0.
\end{gather*}
Substituting (\ref{1parintcond1a}) into the expression (\ref{1-parpoteqns}) for $B^{22}$ we f\/ind
\begin{gather*}
B^{22}=\frac{\big(\frac{\mu_1}{\mu}+\frac{\mu_2}{\mu}B^1\big)\big(\big(B^1\big)^2-1\big)-\big(B_2^1+B^1B^1_1\big)}{B^1}.
\end{gather*}
and, as before,
\[B^{12}=-\frac{\mu_2}{\mu}B^1-\frac{\mu_1}{\mu}.\]

A key equation for this approach is the integrability condition (\ref{newintcond}) of Section \ref{section3}, derived from consideration of $\Lambda_{12}=0$:
\begin{gather}\label{newintconda} 5L^{(1)}\mu_1-5L^{(2)}\mu_2+\big(L^{(1)}_1-L^{(2)}_2+3A_1L^{(1)}-3A_2L^{(2)}\big)\mu=0.
\end{gather}
Recall that $L^{(1)}=A_{112}-A_{12}A_1$, $L^{(2)}=A_{122}-A_{12}A_2$. Here $L^{(1)}=L^{(2)}=0$ is the condition that the system admits a nondegenerate potential. We have already determined  all such 1-parameter systems: they are just the restrictions of nondegenerate potential systems to a single parameter. Our interest is in f\/inding systems that are not simply restrictions, if such systems exist. Thus we require that the coef\/f\/icients of $\mu_1$, $\mu_2$, $\mu$ in (\ref{newintconda}) are all nonvanishing. (Since~(\ref{newintconda}) must admit  a 2-dimensional solution space, if any one coef\/f\/icient vanishes then all vanish.) Setting $B=B^1$ for short, we rewrite conditions~(\ref{1parintcond1a}) and (\ref{newintconda}) in the form
\begin{gather}
\mu_1-B\mu_2+\frac{B_2+BB_1-3B(A_2-BA_1)}{B^2+1}\mu=0,\nonumber
\\
\label{Killing2}
\mu_1-\frac{L^{(2)}}{L^{(1)}}\mu_2+\frac{L^{(1)}_1-L^{(2)}_2+3A_1L^{(1)}-3A_2L^{(2)}}{5L^{(1)}}\mu=0.
\end{gather}
Since there are solutions with arbitrarily chosen values of $\mu_2,\mu$ at a point, these equations must be identical: $\mu_1-B\mu_2+D\mu=0$, where
\begin{gather}\label{Killingidenta} B=\frac{L^{(2)}}{L^{(1)}},\\ \label{Killingidentb}\frac{B_2+BB_1-3B(A_2-BA_1)}{B^2+1}=\frac{L^{(1)}_1-L^{(2)}_2+3A_1L^{(1)}-3A_2L^{(2)}}{5L^{(1)}}=D.
\end{gather}
 Use of $\mu_{12}=0$  and dif\/ferentiation yields the linear expressions
\begin{gather*}
\mu_1=B\mu_2-{D}\mu,\qquad  \mu_{11}=(B_1-DB)\mu_2+(D^2-D_1)\mu,\qquad \mu_{22}=\frac{D_2}{B}\mu+\frac{D-B_2}{B}\mu_2,
\end{gather*}
for $\mu_1$, $\mu_{11}$, $\mu_{22}$ in terms of $\mu$ and $\mu_2$.
Just as in the preceding section, there are additional integrability and compatibility conditions for this system  that constrain $A$ and $B$. The only nontrivial integrability condition for the subsystem $\mu_{12}=0$, $\mu_1-B\mu_2+ D\mu=0$, is
\[ \left[\left(\frac {D_2}{B}\right)_1-\frac{DD_2}{B}\right]\mu+\left[D_2+\left(\frac{D-B_2}{B}\right)_1\right]\mu_2=0.\]
Since this must hold for all solutions $\mu$ we have the 2 conditions
\begin{gather*}
\left(\frac {D_2}{B}\right)_1-\frac{DD_2}{B}=0,\qquad D_2+\left(\frac{D-B_2}{B}\right)_1=0.
\end{gather*}
The requirement that the subsystem be compatible with $\Lambda=0$ is
\[
\left(\frac{D_2}{B}-D^2+D_1+3A_1D-C\right)\mu+\left(\frac{D-B_2}{B}+DB-B_1-3A_1B+3A_2\right)\mu_2=0,\]
i.e.,
\begin{gather*}
\frac{D_2}{B}-D^2+D_1+3A_1D-C=0,\qquad  \frac{D-B_2}{B}+DB-B_1-3A_1B+3A_2=0,
\end{gather*}
where $C=A_{11}+A_1^2-A_{22}-A_2^2$. Note that the second condition is already satisf\/ied.

Since we no longer assume a 4-dimensional space of symmetries, we can no longer deduce directly that $D=-2(B_2+B_1B)/(1+B^2)$. However we shall see that the assumption $L^{(1)}L^{(2)}\ne 0$ and identities (\ref{Killingidenta}), (\ref{Killingidentb}) imply this result.

Indeed we can substitute $B=L^{(2)}/L^{(1)}$ into all identities and express them in terms of $A=\ln A^{12}$ and its derivatives. The verif\/ication that
\begin{gather}\label{required identity} -2(B_2+B_1B)/\big(1+B^2\big)=\frac{L^{(1)}_1-L^{(2)}_2+3A_1L^{(1)}-3A_2L^{(2)}}{5L^{(1)}}=D
\end{gather}
is cumbersome, but straightforward. One uses the identities (\ref{Killingidentb}) and $\Delta=0$ (and its derivatives) to express $A_{1222}$, $A_{1112}$, $A_{122}$, in terms of derivatives of $A$ of lower order (or in some cases equal order with the number of $y$ derivatives $\le$ the number of $x$ derivatives)  and substitutes into both sides of the desired identity until equality is evident. Indeed the only fourth order derivatives occurring in (\ref{required identity}) are $A_{1222}$ and $A_{1112}$. We can solve for these derivatives in terms of strictly lower order derivatives of $A$ from the identities (\ref{Killingidentb}) and $\Delta_{12}=0$. Then we substitute these expressions back into the left and right sides of (\ref{required identity}) to obtain lower order expressions. Similarly we can solve for $A_{122}$ from $\Delta_1=0$ and $A_{22}$ from $\Delta=0$ and substitute back. After several iterations the identity is verif\/ied.

Now we are in exactly the same situation as for the 4-dimensional symmetry case in the previous section. Since
$D=-2(B_2+B_1B)/(1+B^2)$ the system must admit a Killing vector ${\cal X}=Qp_1-QBp_2$. The integrability condition for $Q$ is $D_2=2B_{11}+B_1D+BD_1$ which follows from~(\ref{Killing2}) and other identities listed above.
Thus Lemma \ref{Bfactors} is valid for this case, so $B=L^{(2)}/L^{(1)}=X(x)Y(y)$, i.e., $B$ factors, and the proof of Theorem \ref{L1L2=0} is valid. Thus $L^{(1)}=L^{(2)}=0$, which is a contradiction. Thus  the only 1-parameter superintegrable systems with exactly three second order linearly independent symmetries are those for which $L^{(1)}=L^{(2)}=0$.

\begin{theorem}\label{L1L2=03D} If $2D$ superintegrable system with a $1$-parameter potential $\alpha V$ admits exactly $3$~li\-near\-ly independent second order symmetries then there exists a superintegrable system with nondegenerate potential $\tilde V(\alpha,\beta,\gamma)$ such that the restriction  $\tilde V(\alpha,0,0)=\alpha V$ and  the restricted second order symmetries of the nondegenerate system agree with a three-dimensional subspace of the second order symmetries for the $1$-parameter potential.
\end{theorem}

\section{Structure of systems with 4 linearly independent\\
 second order symmetries}

We begin by proving a fundamental duality result for 2D systems with 4 linearly independent second order symmetries and a 1-parameter potential. (In fact, however, it is not hard to show that every system with 4 independent second order symmetries must admit a Killing vector and a 1-parameter potential.) We have worked out the fundamental structure equations for these systems in Section~\ref{1parsection}. In the generic case they are
\begin{gather}\label{fundintcond20} \mu_{12}=0,\qquad \! a^{12}(\mu_{11}-\mu_{22})+3\mu_1a^{12}_1-3\mu_2a^{12}_2+\big(a^{12}_{11}-a^{12}_{22}\big)\mu=0,\qquad \! a^{12}_{11}+a^{12}_{22}=0,\!\!\!
\\
\mu_1-B\mu_2+ D\mu=0,\qquad D=-2\frac{B_2+BB_1}{B^2+1}=-2\frac{B(a^{12}_2-Ba^{12}_1)}{a^{12}(B^2+1)},\nonumber
\end{gather}
where $V_1=BV_2$ and $B$ satisf\/ies the integrability conditions
\begin{gather}\label{fundintcond22} \big(B^2+1\big)(B_{11}+B_{22})-2B\big(B_1^2+B_2^2\big)=0,\qquad BB_{12}=B_1B_2.
\end{gather}
Using one of the alternate expressions for $D$, we can write the f\/irst equation (\ref{fundintcond20}) in the more obviously dual form
\begin{gather}\label{fundintcond23} \big(B^2+1\big)(\mu_1-B\mu_2)a^{12}+2B\big(Ba^{12}_1-a^{12}_2\big)\mu=0.
\end{gather}
(In the form (\ref{fundintcond20}), (\ref{fundintcond22}), (\ref{fundintcond23}) these equations apply not only to characterize the generic case but also the special cases $B=0,-i$.) The interpretation of these equations is that $\mu=\lambda\ne 0$ is a~metric of a superintegrable system corresponding to the potential function $B$ provided $\lambda$ and $B$ satisfy equations (\ref{fundintcond20}), (\ref{fundintcond22}), (\ref{fundintcond23})  for some 2-dimensional space of harmonic functions $a^{12}$. Then the solution space of all simultaneous solutions $\mu$ is also 2-dimensional and the solutions give precisely the metrics of systems  St\"ackel equivalent to $\lambda$.

The duality result is the following:
\begin{theorem}\label{duality} If $\mu(x,y)$, $a^{12}(x,y)$, $B(x,y)$ satisfy \eqref{fundintcond20}, \eqref{fundintcond22}, \eqref{fundintcond23} then
\begin{gather*}
{\tilde\mu(x,y)}=a^{12}\left(\frac{x+iy}{\sqrt{2}},  \frac{-ix-y}{\sqrt{2}}\right),\qquad
{\tilde a}^{12}(x,y)=\mu\left(\frac{x+iy}{\sqrt{2}},
\frac{-ix-y}{\sqrt{2}}\right),\\
{\tilde B}(x,y)=i\left(\frac{B\left(\frac{x+iy}{\sqrt{2}},  \frac{-ix-y}{\sqrt{2}}\right)-i}{B\left(\frac{x+iy}{\sqrt{2}} \frac{-ix-y}{\sqrt{2}}\right)+i}\right),
\end{gather*}
also satisfy these conditions. Thus, the roles of metric and symmetry can be interchanged if the potential function undergoes an appropriate M\"obius transformation. A second interchange returns the system to its original state.
\end{theorem}

\begin{proof}
This is a simple consequence of the chain rule and the relations
\begin{gather*}
a_1^{12}=\tfrac{1}{\sqrt{2}}({\tilde \mu}_1-i{\tilde\mu}_2),\qquad a_2^{12}=\tfrac{1}{\sqrt{2}}({i\tilde \mu}_1-{\tilde\mu}_2), \\
\mu_1=\tfrac{1}{\sqrt{2}}\big({\tilde a}^{12}_1-i{\tilde a}^{12}_2\big),\qquad \mu_2=\tfrac{1}{\sqrt{2}}\big({i\tilde a}^{12}_1-{\tilde a}^{12}_2\big),
\end{gather*}
with analogous relations between $B$ and $\tilde B$.  Under the duality the f\/irst condition (\ref{fundintcond22}) obeyed by~$B$ maps to the second condition (\ref{fundintcond22}) obeyed by~$\tilde B$, and vice-versa. The theorem also applies to the special cases where $B=0$ or $B=-i$. Under the duality the special cases switch roles.\end{proof}

\begin{theorem} Every $1$-parameter  superintegrable $2D$ system with $4$ linearly independent second order symmetries is
  St\"ackel equivalent to a~superintegrable system on a
  constant curvature space.
\end{theorem}
\begin{proof} We consider the generic case $B\ne 0,-i$ f\/irst. Then  every such 1-parameter  superintegrable 2D system with 4 linearly independent second order symmetries and  metric $\mu=\lambda$  corresponds to the system of equations
(\ref{fundintcond20}), (\ref{fundintcond22}), (\ref{fundintcond23}) where the $a^{12}$ range over a 2-dimensional space and there is a 2-dimensional space of solutions $\mu$, corresponding to the metrics of systems St\"ackel equivalent to the $\mu=\lambda$ system.
 Moreover there is a Killing vector ${\cal X}=\xi p_1+\eta p_2$ such that $a_0^{12}=\xi\eta$ from the second order symmetry ${\cal X}^2$ is nonzero and satisf\/ies the Liouville equation. It follows that
  the dual metric $\xi = \tilde a_0^{12}$ is of constant
  curvature.  The dual system also describes a St\"ackel equivalence class of 1-parameter  superintegrable 2D systems with 4 linearly independent second order symmetries, so there must exist a corresponding Killing vector ${\cal Y}$ and nonzero function $b_0^{12}$ that is harmonic and satisf\/ies the Liouville equation. Since the duality maps all solutions $\mu$ of the original system one to one onto symmetries $b^{12}$ of the dual system,
 there must exist a unique solution
$\mu=\nu$
of the original system  such that $\tilde \nu=b_0^{12}$. Since $b_0^{12}$ satisf\/ies the Liouville equation  $\nu$ is the metric of a constant curvature
space. This means that the system with metric $\lambda$ is St\"ackel
equivalent to the constant curvature system with metric $\nu$.

Now we consider the special cases. If $B=-i$ then all metrics $\mu$ are f\/lat, so the statement of the theorem again holds. If $B=0$ then ${\cal X}=p_1$ so the associated $a^{12}_0=0$. In this case, for any nonzero harmonic $a^{12}$ from the 2-dimensional space of symmetries of the original system we have that ${\tilde a}^{12}$ is a metric for the dual system with potential function ${\tilde B}=-i$. Now such metrics are f\/lat, and as we have shown in the proof of Theorem~\ref{theorem2} there is then a Killing vector $\cal X$ such that the symmetry $b^{12}_0$ from    ${\cal X}^2$ is nonzero and satisf\/ies the Liouville equation with $c=0$. Since the duality map is one to one and onto there must exist a metric $\nu$ in the original system such that $\tilde \nu=b_0^{12}$. This means that $\nu$ is a f\/lat space metric.   \end{proof}

We continue to require that  the space of second order symmetries is of dimension~4
 and now  investigate the space of third order constants of the motion for this system:
\begin{gather*}
{\cal K}=\sum ^2_{k,j,i=1}a^{kji}(x_1,x_2)p_kp_jp_i+b^\ell(x_1,x_2)p_\ell,
\end{gather*}
which must satisfy
$
\{{\cal H},{\cal K}\}=0$.
Here $a^{kji}$ is symmetric in the indices $k$, $j$, $i$.
\begin{theorem}\label{theorem6}  For a $2D$ superintegrable system with $1$-parameter potential and a $4$-dimensional space of second order constants of the motion, the dimension of the space of strictly $3$rd order constants of the motion is at most~$4$.
\end{theorem}

\begin{proof} We give the essence of the proof, leaving out some of the details.
The conditions on $\cal K$  are
\begin{gather}
2 a^{iii}_i=-3((\ln\lambda)_i a^{iii}+(\ln\lambda)_ja^{jii}),\qquad  i\ne j,\nonumber\\
3 a^{jii}_i+
a^{iii}_j=3-((\ln\lambda)_i a^{iij}+(\ln\lambda)_j a^{ijj})        ,\qquad i\ne j,\nonumber\\
2(a^{122}_1+
a^{112}_2)=-(\ln\lambda)_1a^{122}-(\ln\lambda)_1a^{111}
-(\ln\lambda)_2 a^{222}-(\ln\lambda)_2 a^{112},\nonumber
\\
\label{vectpot1}
b^1_2+b^2_1  = 3 \sum_{s=1}^2\lambda a^{s21}
V_s,\quad b^j_j=\frac32 \sum_{s=1}^2a^{sjj}
V_s-\frac12\sum_{s=1}^2(\ln\lambda)_s b^s,
 \qquad j=1,2, \end{gather}
and
\begin{gather}\label{scalpot11}\sum_{s=1}^2b^{s}
V_s=0.
\end{gather}
The $a^{kji}$ is just a third order Killing tensor.
The     $b^\ell$ must depend on the parameter in the potential $V$ linearly, so we have
\[b^\ell(x_1,x_2)=\sum_{j=1}^2f^j(x_1,x_2) V_j(x_1,x_2).
\]
Using the standard notation $V_1=B^1V_2$, $V_{12}=B^{12}V_2$, $V^{11}=B^{11}V_2$, $V^{22}=B^{22}V_2$ with $B^1=B$, we
f\/ind from (\ref{scalpot11}) that $Bf^1\equiv f$, $f^2=-Bf$

Further
\begin{gather*}
 b^1_1=f_1V_2+fV_{12},\qquad b^1_2=f_2V_2+fV_{22},\\
 b^2_1=-(Bf)_1V_2-BfV_{12},\qquad  b^2_2=-(Bf)_2V_2-BfV_{22}.
\end{gather*}
Thus
\begin{gather*}
b^1_1=\big(f_1+fB^{12}\big)V_2,\qquad b^1_2=\big(f_2+f^{22}\big)V_2,\\
 b^1_1=-\big( Bf_1+\big(B_1+BB^{12}\big)f\big)V_2, \quad b^2_2=-\big( Bf_2+\big(B_2+BB^{22}\big)f\big)V_2,
\end{gather*}
Now we assume $B\ne 0$ and $(1+2\lambda)(1-2\lambda+4\lambda^2)\ne 0$. (The last inequality can always be achieved by a rescaling of the coordinates $x_1$, $x_2$, if necessary. We will treat the special case $B\equiv 0$ separately.)
Substituting these results  into the
def\/ining equations (\ref{vectpot1}) and solving for~$f_1$,~$f_2$ and $a^{122}$, we obtain
\begin{gather}\label{feqn1} f_1=\frac32 Ba^{111}+\frac32 a^{112}+\cdots,\\ f_2=\frac{3B\lambda}{1+2\lambda}a^{112}-\frac{3\lambda}{1+2\lambda}a^{222}+\cdots,\label{feqn2}\\
a^{122}=\frac{2B\lambda}{1+2\lambda}a^{112}-\frac{1}{B(1+2\lambda)}a^{222}+\cdots,\label{feqn3}
\end{gather}
where the omitted terms are linear in $f$.
Substituting these results into the remaining 5 conditions, we can recast them in the form
\begin{gather} \label{aeqn1}a^{111}_1=\cdots,\\
a^{111}_2=-3s+\cdots,\label{aeqn2}\\
a^{222}_1=-\frac{6}{1-2\lambda+4\lambda^2}s+\cdots,\label{aeqn3}\\
a^{222}_2=\cdots,\label{aeqn4}\\
a^{112}_1=s,\label{aeqn5}\\
a^{112}_2=-\frac{1+2\lambda}{B(1-2\lambda+4\lambda^2)}s+\cdots,\label{aeqn6}
\end{gather}
where the omitted terms are linear in $f$, $a^{111}$, $a^{222}$, $a^{112}$. (Here we have introduced a new variable $s=a^{112}_1$.) The system
(\ref{feqn1})--(\ref{aeqn6}) is not in involution because we have not expressed the derivatives $s_1$, $s_2$ in terms of $f$, $a^{111}$, $a^{222}$, $a^{122}$, $s$. Dif\/ferentiating (\ref{aeqn2}) with respect to $x_1$,  using the fact that $(a^{111}_2)_1=(a^{111}_1)_2$, and dif\/ferentiating (\ref{aeqn1}) with respect to $x_2$, we can solve for $s_1$. Similarly, dif\/ferentiating (\ref{aeqn3}) with respect to $x_2$,  using the fact that $(a^{222}_1)_2=(a^{222}_2)_1$, and dif\/ferentiating (\ref{aeqn4}) with respect to $x_1$, we can solve for $s_2$. Thus we obtain
\begin{gather} \label{seqn1}s_1=\cdots,\\
s_2=-3s+\cdots,\label{seqn2}
\end{gather}
where the omitted terms are linear in $ f$, $a^{111}$, $a^{222}$, $a^{112}$. We conclude that the full system with 5 functions $s$, $f$, $a^{111}$, $a^{222}$, $a^{112}$ is in involution. Since the space of Killing vectors is one-dimensional we can always add a multiple $\alpha {\cal X}$ to any 3rd order symmetry (where $\alpha$ is the parameter in the potential) and obtain another 3rd order symmetry. Thus, by adding such a~multiple,  we can prescribe  $f=0$, $s$, $a^{111}$, $a^{222}$, $a^{112}$ at a regular point and there is at most one strictly  3rd order symmetry taking on these values  at this point. Hence the space of 3rd order symmetries is at most 4-dimensional.

In the special case $B=0$ we f\/ind $b^1=fV_2$, $b^2=\lambda_1=0$, so $a^{222}=0$. A similar argument to the preceding shows that the system with 5 functions $s$, $f$, $a^{111}$, $a^{112}$, $a^{122}$ is in involution.   Again the space of strictly 3rd order symmetries is at most 4-dimensional.  \end{proof}

Theorem \ref{theorem6} gives important information about the structure of the Poisson algebra generated by the f\/irst and second order symmetries of this superintegrable system  As we know, a basis for the second order symmetries can be taken in the form ${\cal L}_0 ={\cal H}$, ${\cal L}_1$, ${\cal L}_2$, ${\cal X}^2$, where $\cal X$ is the Killing vector. Clearly the set of 4 third order symmetries ${\cal L}_0{\cal X}$, ${\cal L}_1{\cal X}$, ${\cal L}_2{\cal X}$, ${\cal X}^3$ is linearly independent. Hence it must be a basis.

\begin{corollary} \label{corollary6}  For a $2D$ superintegrable system with $1$-parameter potential and a $4$-dimensional space of second order constants of the motion, the dimension of the space of $3$rd order constants of the motion is exactly~$4$.
\end{corollary}

From this result we see that the algebra generated by ${\cal L}_0$, ${\cal L}_1$, ${\cal L}_2$, ${\cal X}$ is closed under the Poisson bracket operation.  Indeed the Poisson brackets $\{{\cal X},{\cal L}_j\}$ are second order symmetries, so we have
\[\{{\cal X},{\cal L}_j\}=a_{j0}{\cal L}_0+a_{j1}{\cal L}_1+a_{j2}{\cal L}_2+a_{j3}{\cal X}^2, \]
where the $a_{jk}$ are constants. The Poisson brackets $\{{\cal L}_i,{\cal L}_j\}$ are third order symmetries, so by Corollary \ref{corollary6} they can be expanded in the form
\[\{{\cal L}_i,{\cal L}_j\}=b_{ij,0}{\cal L}_0{\cal X} + b_{ij,1}{\cal L}_1{\cal X} +b_{ij,2}{\cal L}_2{\cal X} +b_{ij,3}{\cal X}^3.\]
Additional commutators and relations between the constants follow easily from the Jacobi identity.

This is not the full story for these Poisson algebras, because the four linearly independent generators ${\cal L}_0$, ${\cal L}_1$, ${\cal L}_2$, ${\cal X}^2$ must be functionally dependent. We need to understand the form of this dependence. The key to this is the space of 4th order symmetries.

Next  we investigate the space of fourth order constants of the motion.
 Here a constant of the motion
\begin{gather*}
{\cal F}=\sum ^2_{\ell,k,j,i=1}a^{\ell kji}(x,y,z)p_\ell p_kp_jp_i+\sum ^2_{m,q=1}b^{mq}(x,y,z)p_mp_q+W(x,y,z),
\end{gather*}
 must satisfy
$
\{{\cal H},{\cal F}\}=0$.
 Again $a^{\ell  kji}$, $b^{mq}$ are symmetric in all indices.

The conditions are
\begin{gather}\label{4rthordersymalg1}  a^{iiii}_i=-2\sum_{s=1}^2a^{siii}(\ln\lambda)_s,\\
\label{4rthordersymalg2}  4a^{jiii}_i+a^{iiii}_j=-6\sum_{s=1}^2a^{siij}(\ln\lambda)_s,\qquad i\ne j,\\
\label{4rthordersymalg3}  3a^{jjii}_i+2a^{iiij}_j=-\sum_{s=1}^2a^{siii}(\ln\lambda)_s-3\sum_{s=1}^2a^{sijj}(\ln\lambda)_s,\qquad i\ne j,
\\
\label{tenspot2}
2b^{ij}_i + b^{ii}_j =6\lambda\sum_{s=1}^2a^{sjii} V_s-\sum_{s=1}^2b^{sj}(\ln\lambda)_s,\qquad i\ne j,
\\
b^{ii}_i =2\lambda\sum_{s=1}^3a^{siii}V_s-\sum_{s=1}^2b^{sj}(\ln\lambda)_s,\nonumber
\end{gather}
and
\begin{gather}\label{scalpot3}W_i=\lambda\sum_{s=1}^2b^{si}V_s.
\end{gather}
Note that the $a^{\ell kji}$ is  a fourth  order Killing tensor. We require the potential $V$ to be superintegrable 1-parameter,  and
 that the highest order terms, the $a^{\ell kji}$ in the constant
 of the motion, be independent of the   parameter in $V$.
The $b^{mq}$ must depend linearly and $W$ quadratically  on this parameter. We ignore the arbitrary additive constant in $W$.

We set
\[ b^{jk}=f^{jk}V_2,\qquad f^{jk}=f^{kj}.
\]
      Then, listing only the f\/irst derivative terms in the $f^{jk}$ explicitly,  conditions
(\ref{tenspot2}) become
\begin{gather}\label{f11,2} 2f^{12}_1+f^{11}_2=\cdots, \\
\label{f22,1} 2f^{12}_2+f^{22}_1=\cdots, \\
 \label{f11,1} f^{11}_1=\cdots,\\
 \label{f22,2} f^{22}_2=\cdots.\end{gather}
From the
integrability condition $\partial_{x_2}W_1=\partial_{x_1}W_2$ for equations (\ref{scalpot3}) we obtain
the condition
\begin{gather}\label{scalpot4} B\lambda_2f^{11}_1+\lambda_2f^{12}_1-B\lambda_1f^{12}_2-\lambda_1f^{22}_2=\cdots
\end{gather}
which does not involve the variables $a^{ijk\ell}$. From conditions (\ref{4rthordersymalg1})--(\ref{4rthordersymalg3}) we obtain
\begin{gather} \label{a1111,1} a^{1111}_1=\cdots, \\
 \label{a2222,2} a^{2222}_2=\cdots, \\
 \label{a1111,2} 4a^{1112}_1+a^{1111}_2=\cdots, \\
 \label{a2222,1} 4a^{1222}_2+a^{2222}_1=\cdots, \\
 \label{a1112,2} 3a^{1122}_1+2a^{1112}_2=\cdots, \\
 \label{a1222,1} 3a^{1122}_2+2a^{1222}_1=\cdots,
 \end{gather}
 where we list only the f\/irst derivative terms in the $a^{ijk\ell}$ explicitly.

\begin{theorem}\label{theorem7}  For a $2D$ superintegrable system with $1$-parameter potential and a $4$-dimensional space of second order constants of the motion, the dimension of the space of strictly $4$th order constants of the motion is at most $9$.
\end{theorem}
\begin{proof} We give the basic ideas, leaving out some of the details. The proof is by parameter counting. There are 8 variables,
\[f^{11}, \ f^{12}, \ f^{22}, \ a^{1111}, \ a^{1112}, \ a^{1122}, \ a^{1222}, \ a^{2222}.
\]
 The number of possible f\/irst derivatives of these variables is 16, and they are subject to the 11 conditions (\ref{f11,2})--(\ref{a1222,1}). The number of possible second derivatives of the variables is 24, and they are subject to the 22 conditions, obtained by dif\/ferentiating the original conditions with respect to $x$ and to $y$, Similarly, the number of possible third derivatives of the variables is~32, and they are subject to the 33 conditions, obtained by taking all second derivatives of the original conditions. At this point the system is in involution, since all third derivatives of the variables can be expressed in  terms of lower order derivatives. The total number of variables is $8+16+24+32=80$ and the number of conditions is $11+22+33=66$, so there are 14 parameters remaining. However, there is one more third order condition than third derivatives, so we can use the 33 conditions to eliminate all of the third derivative terms and obtain a~condition involving at most second derivatives. Explicitly, this
relation is
\begin{gather*}
 \partial_{11}{\bf (\ref{scalpot4})} +\lambda_2\left(-B\partial_{12}+\frac12\partial_{22}\right){\bf (\ref{f11,1})}+\lambda_1\left(\partial_{12}-\frac{B}{2}\partial_{11}\right){\bf (\ref{f22,2})}
\\
\qquad{}-\frac{\lambda_2}{2}\partial_{12}{\bf (\ref{f11,2})}+\frac{B\lambda_1}{2}\partial_{12}{\bf (\ref{f22,1})}=0,
\end{gather*}
where we have indicated the identities to be dif\/ferentiated by their equation numbers. The resulting condition is
\begin{gather}  6B\lambda\lambda_2\left(-Ba^{1111}_{12}+\frac12 a^{1111}_{22}\right)+56\lambda\lambda_1\left(-\frac{B}{2}a^{2222}_{11}+a^{2222}_{12}\right)
+3B\lambda(-\lambda_2+B\lambda_1)a^{1122}_{12}
\nonumber\\
\qquad{}+\lambda\lambda_2\big(-Ba^{1112}_{12}+a^{1112}_{22}\big)+\lambda\lambda_1
\big(-a^{1222}_{11}+Ba^{1222}_{12}\big)=\cdots,\label{newcond}
\end{gather}
where we have listed explicitly only the second derivative terms in the $a$-variables, not the $f$-va\-riables.  It is straightforward to verify that the second order part of condition (\ref{newcond}) is nonzero and that it is not obtainable by taking linear combinations of f\/irst derivatives of conditions (\ref{a1111,1})--(\ref{a1222,1}). Thus it is an independent second order condition. Dif\/ferentiating this condition with respect to $x$ we get a third order condition with leading terms
\begin{gather*}  6B\lambda\lambda_2\left(-Ba^{1111}_{112}+\frac12 a^{1111}_{122}\right)+56\lambda\lambda_1\left(-\frac{B}{2}a^{2222}_{111}+a^{2222}_{112}\right)
+3B\lambda(-\lambda_2+B\lambda_1)a^{1122}_{112}
\nonumber\\ \qquad{}+\lambda\lambda_2\big(-Ba^{1112}_{112}+a^{1112}_{122}\big)
+\lambda\lambda_1\big(-a^{1222}_{111}+Ba^{1222}_{112}\big)=\cdots,
\end{gather*}
which is not obtainable by taking linear combinations of second derivatives of conditions (\ref{a1111,1})--(\ref{a1222,1}). Thus the maximum number of parameters for this system is reduced to 13. However we can add $\alpha^2$ times any linear combination of a basis of four 2nd order symmetries to a 4th order symmetry (where $\alpha$ is the parameter for the potential $V$). Thus the maximum dimension for the space of strictly 4th order symmetries is 9.   \end{proof}

As we pointed out earlier, the basis of 4 second order symmetries
${\cal L}_0$, ${\cal L}_1$, ${\cal L}_2$, ${\cal L}_3={\cal X}^2$
must be functionally dependent. This dependence must occur at the fourth order. Indeed we can form 10 products $\{{\cal L}_j{\cal L}_k : j\le k\}$, from these symmetries, all strictly fourth order. Since the maximum dimension of the space of fourth order symmetries is $\le 9$ it follows that the set $\{{\cal L}_j{\cal L}_k \}$ is linearly dependent. There can only be one such linear dependency relation, and we call it the Casimir. This relation, and the second order and third order structure relations given previously, completely determine the structure of the Poisson algebra.

\section{Discussion and outlook}
We have f\/illed in a gap in the literature by clarifying the structures of the quadratic algebras corresponding to 2D second order superintegrable systems with 1-parameter potentials. For such a system the number of linearly independent second order symmetries is either 3 or 4. If it is 3 the system is simply a restriction from a nondegenerate potential system. It is 4 if and only if it admits a f\/irst order symmetry. For 4 second order symmetries the quadratic algebra is closed at orders three and four from dimensional considerations: The number of terms obtainable from the basis f\/irst and second order symmetries exceeds the maximal dimension of the space of third order and of fourth order symmetries. Every such system is St\"ackel equivalent to a system on a constant curvature space. In 3D and higher dimensions the relation between systems corresponding to degenerate and nondegenerate potentials is much more complicated~\cite{KKM2007, EVA2008, EVA(2)2008} but the tools developed here should be applicable. (Indeed in 3D the potentials can depend on up to 4 parameters, neglecting the additive constant. Nondegenerate (4-parameter) potentials always have 6 linearly independent second order constants of the motion although the number of functionally independent constants is~5. The functional relation between the constants is of order~8 in the momenta, Not all 3-parameter potentials are restrictions of nondegenerate potentials. Some such systems admit a closed symmetry algebra with some generators of order~4~\cite{EVA2008}. There are up to 8 second constants of the motion for 2-parameter potentials.)
It is also important to extend the structure analysis to superintegrable systems of order greater than two.

Finally the 0-parameter (free space) case deserves more attention, even though it was treated by Koenigs \cite{Koenigs}. If we consider a true 0-parameter potential system as one in which it is not possible to admit a nonconstant potential, then it follows from our results and those of Koenigs that no such system exists. Koenigs' proof used complex variable techniques, very dif\/ferent from the methods used here. Further the last stages of his argument where he shows that no superintegrable systems exist other than those he has found are somewhat murky. Koenigs does not consider potentials or the St\"ackel transform, and
as we have shown, these are very important in the theory. Thus there are good reasons for a new approach to uncovering the structure of these systems.

\pdfbookmark[1]{References}{ref}
\LastPageEnding
\end{document}